\documentclass{elsart}
\usepackage{graphicx}
\usepackage{amsmath}
\usepackage{amssymb}

\newcommand{\hetrois}    {\mbox{$ ^{3}{\mathrm{He}}                            $}~}
\newcommand{\hetro}    {\mbox{$ ^{3}{\mathrm{He}}                            $}}
\newcommand{\tritium}    {\mbox{$ ^{3}{\mathrm{H}}                            $}~}

\newcommand{\neut}{$\tilde{\chi}$~}
\newcommand{\neutt}{$\tilde{\chi}$}

\newcommand{\gam} {{$\gamma$-ray}~} 
\newcommand{\gams} {{$\gamma$-rays}~}

\def\NIMA#1#2#3{{\rm Nucl.~Instr.~and~Meth.} {\bf#1} (19#2) #3}

\def\PRB#1#2#3{{\rm Phys. Rev.} {\bf{B#1}} (19#2) #3}

\def\PRL#1#2#3{{\rm Phys.~Rev.~Lett.} {\bf{#1}} (19#2) #3}
\begin{document}
\runauthor{ }
\begin{frontmatter}
\title{Design optimization of MACHe3,\\ a project of superfuid \hetrois detector for direct Dark Matter search.}
\author{F. Mayet $^{a,}\!$\thanksref{corr}}
\author{, D. Santos $^{a,}\!$\thanksref{corr}}
\author{, G. Perrin $^{a}$,}
\author{ Yu. M. Bunkov $^{b}$, H. Godfrin $^{b}$}
\thanks[corr]{corresponding authors : Frederic.Mayet@isn.in2p3.fr, santos@isn.in2p3.fr, tel: +33 4-76-28-40-21, fax: +33
4-76-28-40-04}
\address{$^{a}$ Institut des Sciences Nucl\'eaires, \\
 CNRS/IN2P3 and Universit\'e Joseph Fourier, \\
 53, avenue des Martyrs, 38026 Grenoble cedex, France}
\address{$^{b}$ Centre de Recherche sur les Tr\`es Basses Temp\'eratures, \\
 CNRS, BP166, 38042 Grenoble cedex 9, France} 
\begin{abstract}
MACHe3 (MAtrix of Cells of superfluid \hetro) is a project of a new detector for direct Dark Matter (DM) search. A cell of superfluid \hetrois
has been developed and the idea of using a large number of such cells in a high granularity detector is proposed.
\noindent
This paper presents, after a brief description of the superfluid \hetrois cell, the simulation of the response 
 of different matrix configurations allowing to define an optimum design as a function of the number of cells and the volume
 of each cell. The background rejection, for several configurations, is presented both for neutrons and \gams of various kinetic
 energies.
\end{abstract}
\end{frontmatter}
\newpage
\section{Introduction to MACHe3}
\begin{figure}
\begin{center}
\includegraphics[scale=0.55]{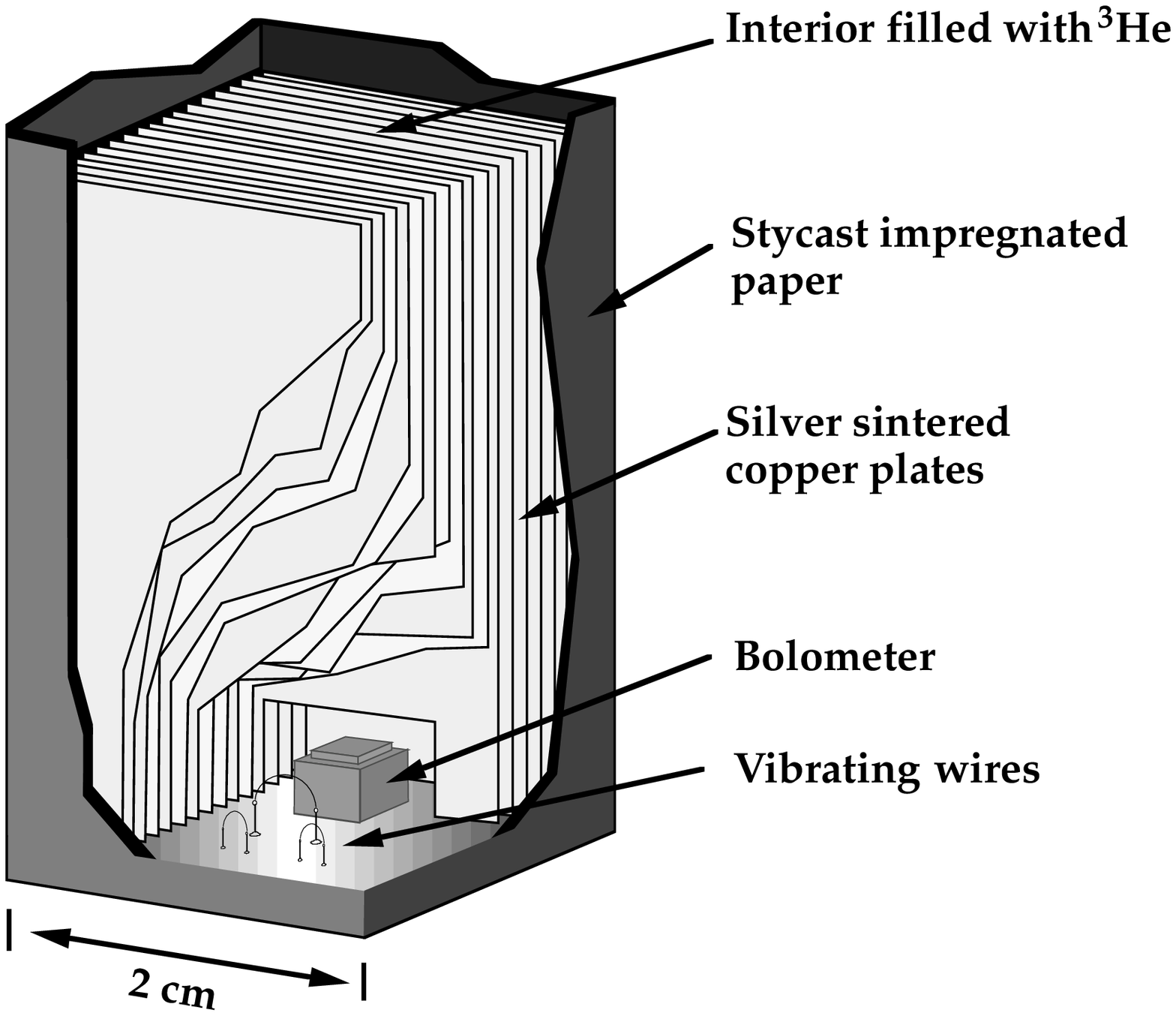}
{\noindent
\caption{Ultra-low temperature nuclear stage and bolometer cell.
\label{fig:design}}}
\end{center}
\end{figure}

\begin{figure}
\begin{center}
\includegraphics[scale=0.55]{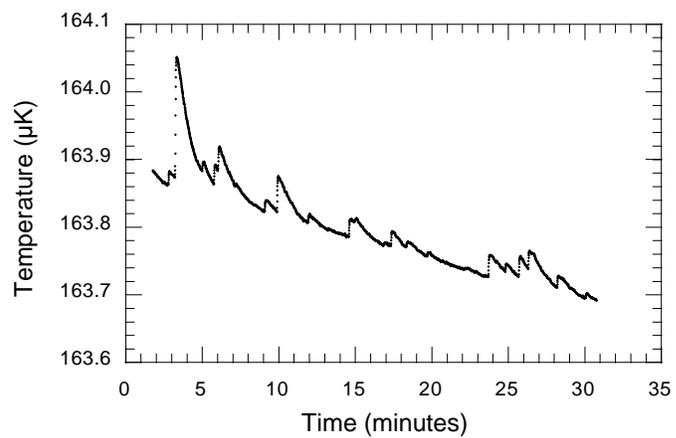}
{\noindent
\caption{Temperature recorded inside the bolometer as a function of time.
\label{fig:signal}}}
\end{center}
\end{figure}

As previously suggested \cite{vieux1,vieux2}, superfluid \hetrois provides a suitable working medium for the detection of low energy 
recoil interactions. Recent studies \cite{prl95} have shown the possibility to use a superfluid \hetrois cell at ultra low temperatures
(T$\simeq$100 $\mu$K).
The primary device consisted of a small copper cubic box (V$\simeq$ 0.125 cm$^{3}$) filled with \hetro. It is immersed in a larger
volume containing liquid \hetrois and thin plates of copper nuclear-cooling refrigerant, see fig. \ref{fig:design}. 
Two vibrating wires are placed inside the cell, forming a Lancaster type bolometer\cite{prl95}. A small hole on one 
of the box walls connects the box to the main \hetrois volume, thus allowing the diffusion of the thermal excitations of the \hetrois generated 
by the energy deposited in the bolometer by the interacting particle.\\
This high sensitivity device is used as follows : the incoming particle deposits an amount of energy in the cell, which is converted into 
\hetrois quasiparticles. These are detected by their damping effect on the vibrating wire. It must be pointed out that the size of the
hole governs the relaxing time (quasiparticles escape time) and the Q factor of the resonator governs the rising time, see figure 
\ref{fig:signal}. The present device has a rather high Q factor (Q $\simeq 10^{4}$), giving a rising time of the order of one second.
Although the primary experiment was still rudimentary, it has allowed to detect signals down to a threshold of 1 keV \cite{prl95}.
Many ideas are under study to improve the sensitivity of such a cell. Recently, the fabrication of 
micromechanical silicon resonators has been reported \cite{trique} and the possibility to use such wires at
ultra-low temperatures is under study.\\
The aim of the present article is to show that, by using a large number of these cells,  a 
high granularity superfluid \hetrois detector could be used for direct Dark Matter (DM) search\footnote{It should be noticed that such a device need 
to be placed in an underground site to 
reduce cosmic rays (mainly muons) background. It should also be surrounded by neutron and \gam shieldings.}. 
For this purpose we have evaluated, by simulation of different kind of background events, the rejection coefficients that may be achieved with such a device.
\section{Particle interactions in \hetrois}
As other direct DM search detectors (Edelweiss\cite{edel}, CRESST\cite{cresst}, CUORE\cite{cuore}), the identification of 
WIMPs (\neutt)\footnote{In particular, we shall suppose all through this work a neutralino (\neutt), the lightest supersymmetric
particle, as the particle making up the bulk of galactic cold DM.} may be obtained by detecting 
their elastic scaterring on a nucleus of a sensitive medium. 
In the case of \hetro, the \neut is expected to transfer \cite{next} up to 6 keV. The maximum \hetrois recoil energy is given by :
\begin{center}
$\mathrm{E}^{max}_{recoil}=2\times \frac{mM^{2}}{(m+M)^2}\times v^2$
\end{center}
\noindent
where $m$ is the mass of the \hetrois nucleus, $M$ the mass of the \neut and $v$ is the relative speed of the \neut.
Assuming that $M \gg m$, as the accelerator experiments claim , this relation yields to $\mathrm{E}^{max}_{recoil}=2mv^2$, which gives for $v \simeq 300 km.s^{-1}$, 
a maximum recoil energy of $\sim$ 6 keV.\\
Hence, in order to evaluate the
expected background for such a detection, it is necessary to know the proportion of events releasing less than 6 keV in the 
\hetrois cell.
The main background components for direct DM search are : thermal and fast neutrons, muons and gamma rays.
\subsection{Neutron interaction in \hetrois}
\label{sec:intera}
The total cross-section interaction for a neutron in \hetrois ranges from $\sigma_{tot} \simeq 1000$ barns, 
for low energy neutrons(E$_{n} \simeq$ 1 eV), down to $\sigma_{tot} \simeq 1$ barn
for 1 MeV neutrons. The main processes are : elastic scattering which starts being predominant above 600 keV, and neutron 
capture: \hetro(n,p)\tritium , which is largely predominant for low energy neutrons ($E_{n} \leq 10$ keV) :
\begin{center}
n+ $^{3}$He $\rightarrow$ p+$^{3}$H +764 keV
\end{center}
\noindent
The energy released by the neutron capture is shared by the recoil ions : the tritium $^{3}$H with kinetic energy 191 keV 
and the proton with kinetic energy 573 keV. The range \cite{ltemp} for these two particles
is fairly short : typically 12 $\mu$m for tritium and 67 $\mu$m for proton; consequently neutrons undergoing capture in \hetrois are expected 
to produce 764 keV within the cell, 
thus being clearly separated from the expected \neut signal (E $\leq $ 6keV). The tritium produced by neutron capture
will eventually decay with a half-life of 12 years by $\beta$-decay with an end-point electron spectrum at 18 keV. It 
means that the number of neutrons capture per cell must be counted to estimate the contribution of this kind of events
on the false \neut rate.\\
The capture cross-section decreases with increasing neutron kinetic energy, but on the other hand, the energy released in 
the \hetrois cell by the elastic scattering is getting larger, thus
diminishing the probability to leave less than 6 keV. From this, it is clear that the worst case will be 8 keV neutrons for which the capture 
process is
less predominant, and the energy left by (n,n) interaction is always less or equal to 6 keV.\\
In order to reduce contamination from neutron background, the idea is either to have a correlation among the cells, which means a large number 
of \hetrois cells, or to have a cell large enough for the neutron to be slowed down until it is captured.
\subsection{$\gamma$-ray interaction in \hetrois}
As \hetrois presents the property to have a low
photoelectric cross-section, Compton scaterring is largely predominant between 100 keV and 10 MeV (for 100 keV \gams : 
$\sigma_{comp}/\sigma_{phot} \simeq$ 10). Consequently, the strategy to
separate a \neut event from a $\gamma$-ray event is two fold : either the cell is large enough for the $\gamma$-ray to undergo multi-Compton 
scattering within one cell \footnote{This will of course be efficient for an energy greater than 10 keV.}, or the number of cells in the 
matrix is large enough so that there could be an
interaction in more than one cell (this will be referred to as a correlated event or a multi-cell event). It will be shown, in the next
section, that having a relatively large cell in a large matrix presents the best rejection against $\gamma$-ray events.\\
The copper used to build the cells must be produced by a controlled procedure with respect to the radioactive contaminations. Nevertheless, the interaction
of \gams with the copper will produce X-rays and scaterred electrons by Compton and photo-electric  interactions. These kind of interactions have been
taken into account  in our simulations and they enhance the correlation among the different cells fired by an incoming \gam.
\section{Simulation of the response of MACHe3 to background events.}
\label{simumatrix}
\begin{figure}
\begin{center}
\includegraphics[scale=0.40]{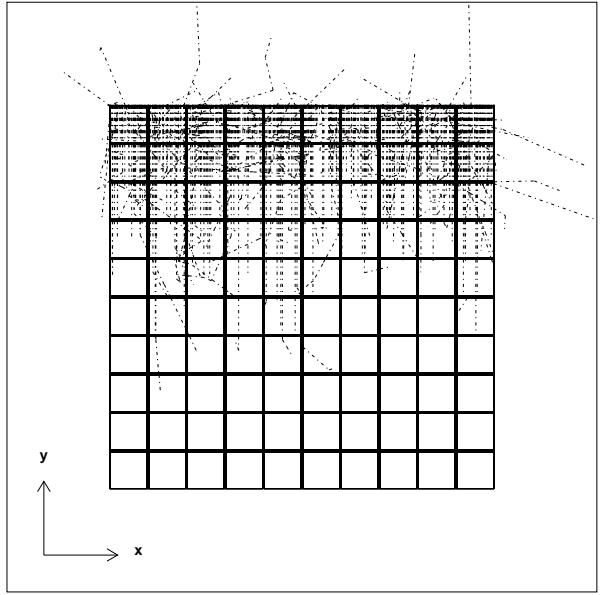}
{\noindent
\caption{2-dimensionnal view of a proposed matrix of 1000 cells (125 $cm^{3}$ each). The events generated in a direction 
perpendicular to the upper face, are 10 keV neutrons. It can be noticed that most of
neutrons of this energy are captured in the first layer.
\label{fig:10kev}}}
\end{center}
\end{figure}

The aim of this simulation is to evaluate the capability of a superfluid \hetrois matrix to reject background events, by taking advantage 
both on correlation among the cells (multi-cell events) and energy loss measurement. The simulation has been done with a complete Monte-Carlo 
simulation using GEANT3.21 \cite{geant} package and in particular the GCALOR-MICAP(1.04/10) \cite{micap} package for slow neutrons.
The simulated detector consists of a cube containing a variable number of cubic \hetrois cells, as it can be seen on figure \ref{fig:10kev}. It is immersed in a large volume containing \hetrois
($\rho_{SF}$=0.08 $g.cm^{-3}$). Each
cell is surrounded by a thin copper layer and it is separated from the others by a gap of 2 $mm$ (filled with \hetro). The events are generated 
in a direction perpendicular\footnote{It has been checked that this procedure does
not affect the values and general behaviour of the matrix parameters, keeping the calculation time short.} to one of the matrix faces. The number of
events per simulation is of the order of 200$\times10^{3}$.
The idea is to find the best matrix design (number of cubic cells and the size of each cell) for which
 the rejection power, taking into account the correlation among the cells and the energy loss measurement, is the highest.  

\noindent
As said previously, a typical \neut is expected to release less than 6 keV in the \hetrois cell. As the elastic 
cross-section between a \neut and \hetrois is fairly small ($\sigma \lesssim
10^{-3} pb$), a
\neut event is expected to be characterized by a single-cell event, with equal probability among all the cells of the matrix.\\
Consequently, the rejection against background events will be achieved by choosing only events having the following characteristics :
\begin{itemize}
{
\item Only one cell fired (single-cell event). The quality parameter related to this selection will be defined below as C$_{geo}$.
\item Energy measurement in this cell below 6 keV and above a threshold of 0.5 keV (quality parameter : R$_{ener}$).
\item An additional constraint can be imposed : the fired cell is in the inner part of the matrix (quality parameter : C$_{veto}$). This condition, which considers the outermost cell layer
as a veto, will allow to reject low energy neutrons interacting elastically, as shown below.
}
\end{itemize}
Let N be the number of events giving a signal in the matrix (any energy, any number of cells), N$_{1}$ the number of single-cell events 
(any energy) and N$_{6}$ the number of  single-cell events with an energy measurement below 6 keV. M$_{1}$ and M$_{6}$ will be referred with
the same meaning as  N$_{1}$ and N$_{6}$, but for events firing a cell in the inner part of the detector (out of the veto).\\
Then, we may define the following parameters as :
\begin{itemize}
{
\item C$_{geo}$$=\frac{\mathrm{N}_{1}}{\mathrm{N}}$ ; the correlation coefficient (proportion of single-cell events).
\item R$_{ener}=\frac{\mathrm{N}_{1}}{\mathrm{N}_{6}}$ ; the rejection by energy measurement.
\item C$_{veto}=\frac{\mathrm{N}_{1}}{\mathrm{M}_{1}}$ ; the veto coefficient.
\item R$_{int}=\frac{\mathrm{N}}{\mathrm{M}_{6}}$ ; the intrinsic rejection.}
\end{itemize}

\subsection{Design optimization.}
In order to define the optimimum matrix design (number of cubic cells and size of the cells), a complete simulation has been done.
The results concerning three types of background are presented : 10 keV
neutrons, 1 MeV neutrons and 2.6 MeV $\gamma$-rays. For each sample, the four parameters defined above are evaluated in various configurations : cell size
of 0.5, 1.0, 2.5 and 5.0 $cm$ and matrix containing $3^{3}$, $5^{3}$, $7^{3}$, $10^{3}$ ($20^{3}$) cells. The best design will be the one for which C$_{geo}$
 is the lowest (thus minimizing the proportion of single-cell background events) and R$_{ener}$ is the highest (meaning a low proportion of background events
  with an energy measurement below 6 keV).
\subsection{$\gamma$-ray background.} 
\begin{figure}
\begin{center}
\includegraphics[scale=0.55]{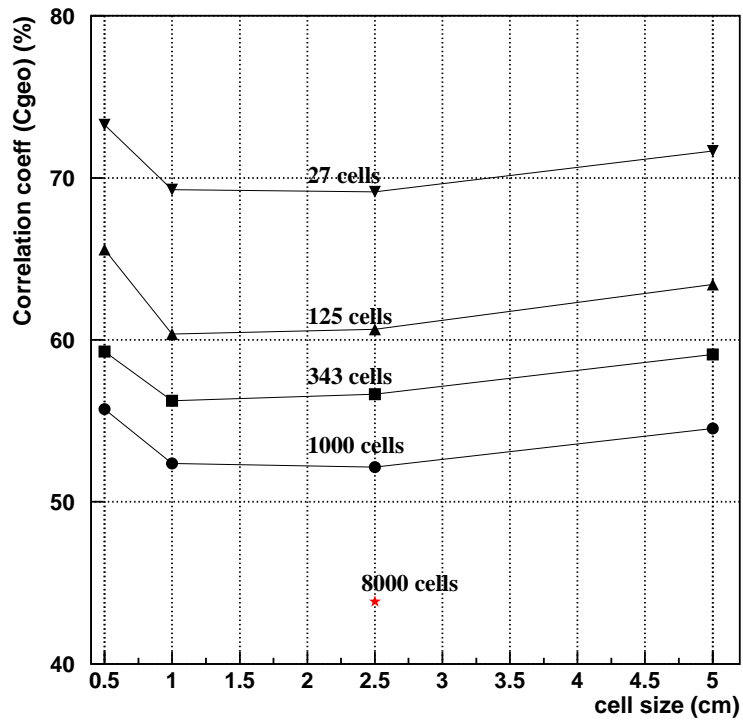}
\caption{Correlation coefficient (C$_{geo}$) as a function of the size of the \hetrois cell for 2.6 MeV $\gamma$-rays. 
The different curves correspond to different matrix sizes as indicated by the labels. 
\label{fig:cgeo26m}}
\end{center}
\end{figure}

Due to the fact that it is a simulation without any constraint on the detector volume, the correlation coefficient depends
strongly on the size of the matrix, with a small dependence on the cell size, as shown on figure \ref{fig:cgeo26m}. 
The best correlation is obtained for 8000 cells of size
2.5 $cm$ (C$_{geo} \simeq 45 \%$). In order to keep a reasonable number of cells, it can be noticed that a matrix 
of same volume (1000 cells of 5.0 $cm$ side) presents also a good correlation 
(C$_{geo} \simeq 55 \%$). A multi-cell event can either be a multi-Compton event, or a single-Compton event for which the 
electron is escaping the cell and firing a neighbouring cell. This last process depends mainly on the cell size and explains the fact that C$_{geo}$ 
remains constant for cell sides larger than 1 $cm$, see fig. \ref{fig:cgeo26m}.\\
It has been found that the energy rejection (R$_{ener}$) depends mainly on the size of the cell.
For a large cell (5 $cm$ side), a rejection R$_{ener} \simeq 90 $ is obtained, allowing to reject 98 \%
of the 2.6 MeV $\gamma$-rays.
The total rejection (see fig. \ref{fig:rejall}), which take into account the correlation and energy selection, together with 
veto selection and interaction probability, is R$\simeq$700 for 1000 cells of size 5.0 $cm$ (for 2.6 MeV \gams).\\
Consequently, for \gam background rejection purpose, a cell of 5.0 $cm$ side presents the best energy rejection and a matrix of 1000 cells 
of this size allows to obtain a good correlation coefficient. In section \ref{gamrejsec}, the rejection of such a matrix as a function
of the \gam energy will be presented.
\subsection{Low energy neutron background.}
\begin{figure}
\begin{center}
\includegraphics[scale=0.55]{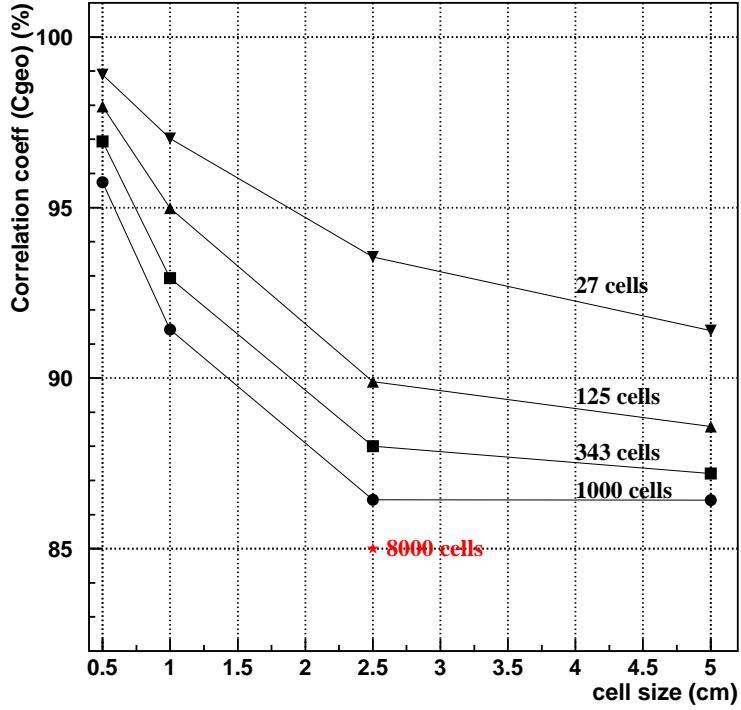}
\caption{Correlation coefficient (C$_{geo}$) as a function of the size of the \hetrois cell for 10 keV neutrons. The different curves
correspond to different matrix sizes as indicated by the labels.
\label{fig:cgeon10}}
\end{center}
\end{figure}

\begin{figure}
\begin{center}
\includegraphics[scale=0.55]{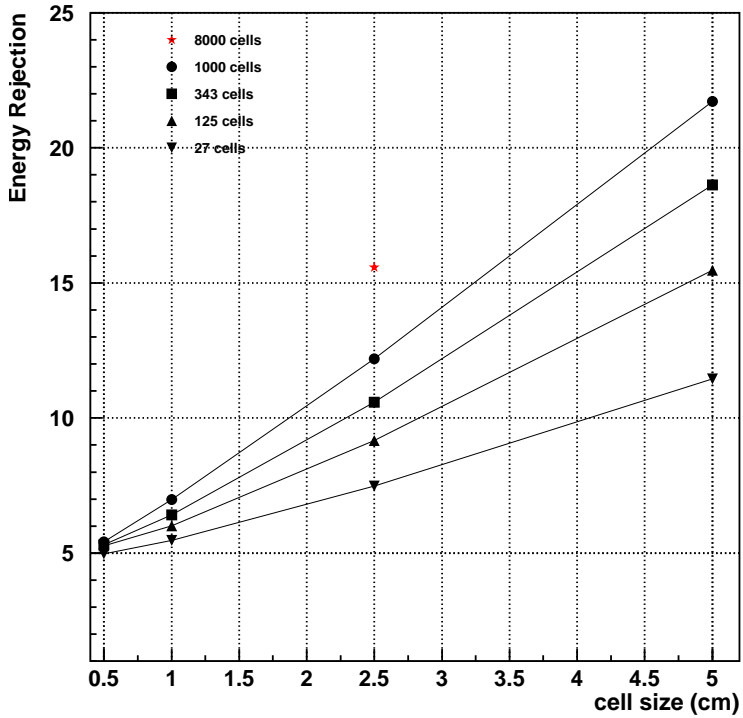}
\caption{Energy Rejection (R$_{ener}$) as a function of the size of the \hetrois cell for 10 keV neutrons. The different curves
correspond to different matrix sizes as indicated by the labels. 
\label{fig:cenern10}}
\end{center}
\end{figure}

\begin{figure}
\begin{center}
\includegraphics[scale=0.8]{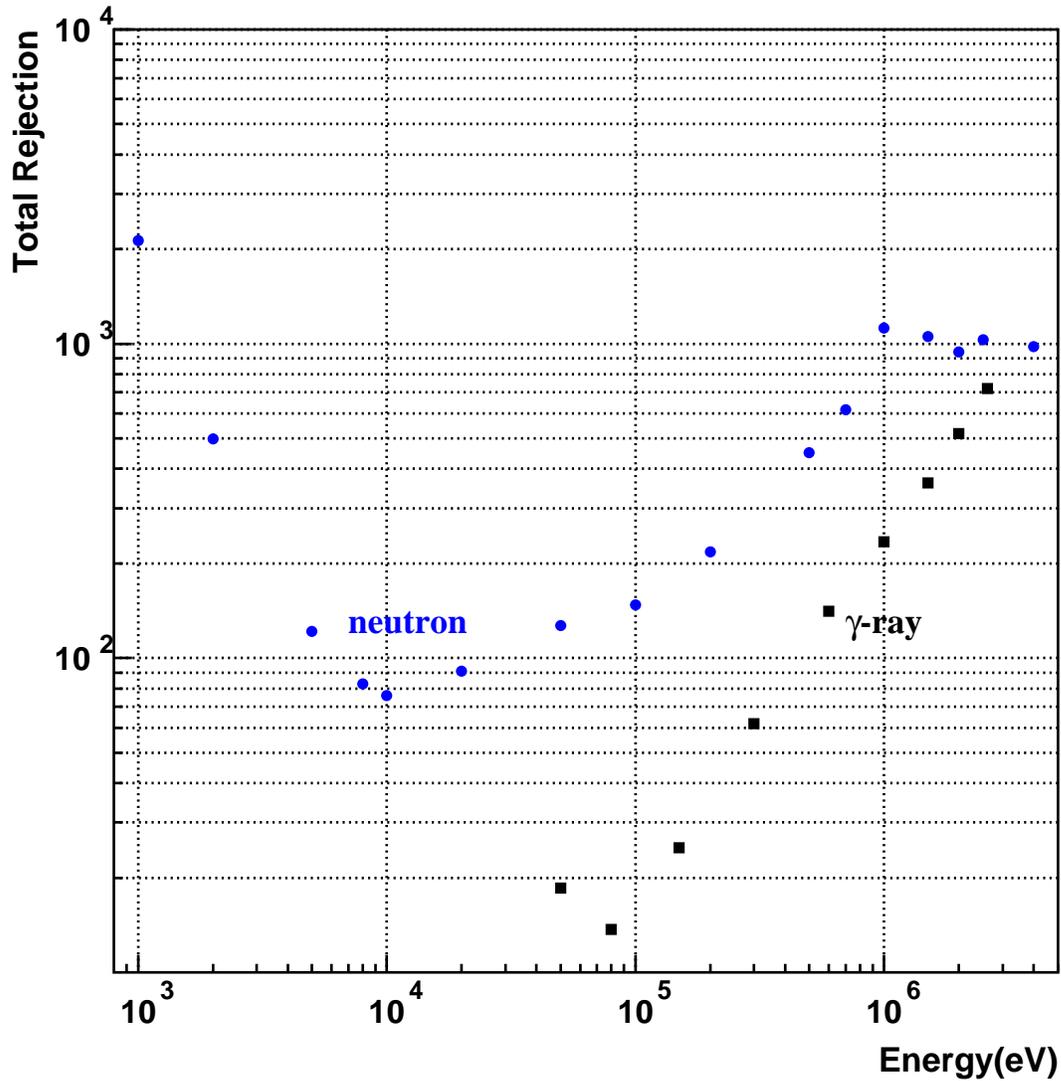}
{\noindent
\caption{Total Rejection as a function of the incident particle energy, for a matrix of 1000 cells (125
$cm^{3}$ each). The different set of points correspond to \gams (squares) and neutrons (circles). The total rejection is defined as the ratio 
between the number of incoming particles and the number of false \neut events (less than 6 keV in one non-peripheric cell).
\label{fig:rejall}}}
\end{center}
\end{figure}

Figures \ref{fig:cgeon10} and \ref{fig:cenern10} present the correlation coefficient (C$_{geo}$) and the energy rejection 
(R$_{ener}$) as a function of the cell size, for different matrix sizes and an incident neutron energy of 10 keV. 
The correlation coefficient depends both on the size of the cell
and of the matrix, since the neutron capture is the predominant process at this energy. 
The best correlation is obtained for 8000 cells of size
2.5 $cm$ (C$_{geo} \simeq 85 \%$), but a larger cell (5 $cm$ side), with only 1000 cells presents also a similar correlation 
(C$_{geo} \simeq 86 \%$). The energy rejection (R$_{ener}$) depends not only on the size of the cell, but also on the size 
of the matrix. 
The best rejection (R$_{ener}\simeq 22$) is achieved for a large cell (5 $cm$ side) and a large matrix (1000 cells). 
The total rejection, shown on fig.\ref{fig:rejall}, is R$\simeq$80 for 1000 cells of 5 $cm$ side, meaning that only 1.25 \% of the incoming
10 keV neutrons may simulate a \neut event. It must be pointed out that 10 keV represent the worst case for rejection purpose, as it can be
seen on figure \ref{fig:rejall}. 

\subsection{Fast neutron background.}
\begin{figure}
\begin{center}
\includegraphics[scale=0.55]{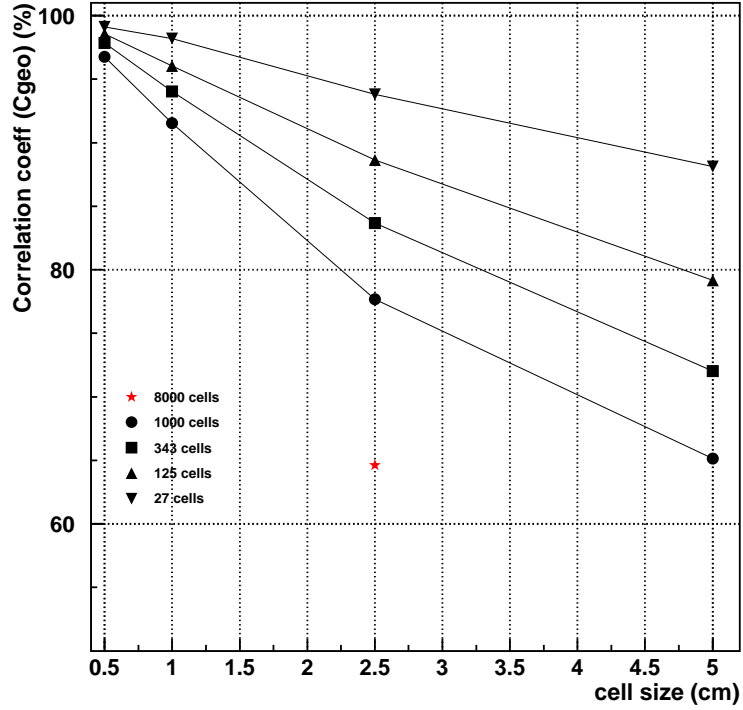}
\caption{Correlation coefficient (C$_{geo}$) as a function of the size of the \hetrois cell for 1 MeV neutrons. The different curves
correspond to different matrix sizes as indicated by the labels. 
\label{fig:cgeon1}}
\end{center}
\end{figure}

\begin{figure}
\begin{center}
\includegraphics[scale=0.55]{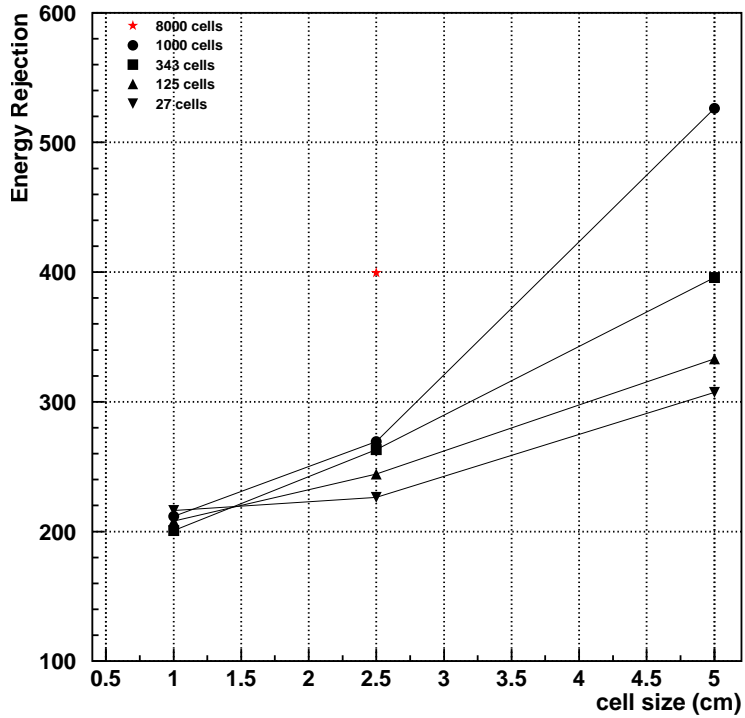}
\caption{Energy Rejection (R$_{ener}$) as a function of the size of the \hetrois cell for 1 MeV neutrons. The different curves
correspond to different matrix sizes as indicated by the labels. 
\label{fig:cenern1}}
\end{center}
\end{figure}

Figures \ref{fig:cgeon1} and \ref{fig:cenern1} show the correlation coefficient (C$_{geo}$) and the energy rejection 
(R$_{ener}$) as a function of the cell size, for different matrix sizes and an incident neutron energy of 1 MeV. As well as 
for low energy neutrons, the
correlation coefficient depends on the matrix and cell sizes. A correlation of $\sim$ 65 \% is achieved for a
large \hetrois volume (1000 cells of 5 $cm$ or 8000 cells of 2.5 $cm$). 
A large matrix of big cells (1000 cells of 5 $cm$) allows to obtain a rather large energy rejection (R$_{ener} \simeq 500$),
leading to a total rejection of the order of 1000 (see fig.\ref{fig:rejall}), meaning that 99.9\% of 1 MeV neutrons arriving on
the \hetrois matrix may be discriminated from a \neut event.\\
\noindent
For these three particle samples, the simulation has shown that a large cell (125 $cm^3$) allows to obtain 
a large energy rejection, and a large matrix (1000 cells or more) allows to have a good correlation among cells, 
thus rejecting efficiently \gams and neutrons of kinetic energy E$\simeq$1 MeV. Hence, for background 
rejection consideration the optimum configuration is a matrix of 1000 cells of 5 $cm$ side.

\subsection{Rejection power of a superfluid \hetrois matrix.}
As shown previously, a matrix of $10^{3}$ large cells (125 $cm^{3}$ each) presents the best rejection power, both for neutrons and
$\gamma$-rays. This section presents the various coefficients, as defined in section \ref{sec:intera}, as a function of the energy of the 
incoming particle.\\
\subsubsection{Rejection against \gam background.}
\label{gamrejsec}
\begin{figure}
\begin{center}
\includegraphics[scale=0.55]{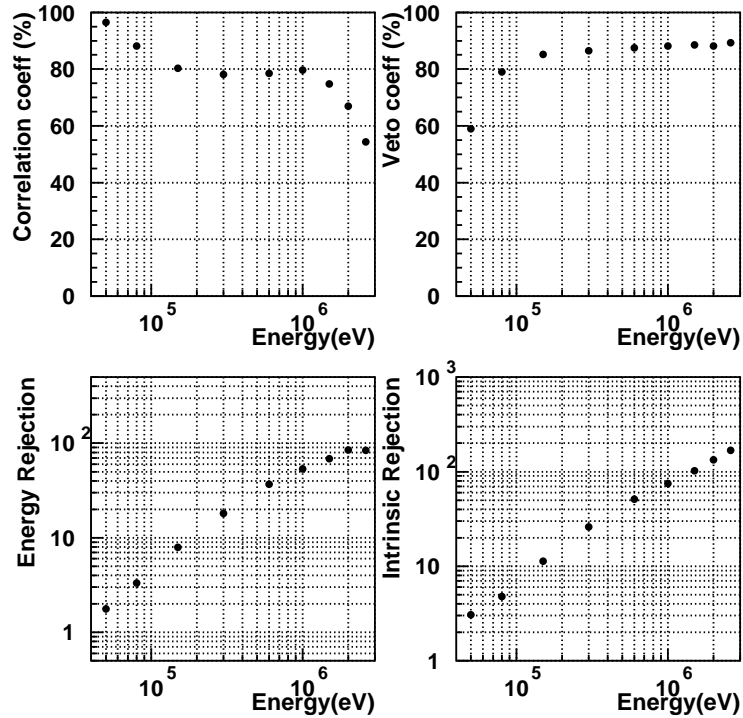}
{\noindent
\caption{\gams interacting in MACHe3 : The four matrix parameters, defined in sec. \ref{simumatrix}, as a function of the $\gamma$-ray energy, for a matrix of 1000 cells (125
$cm^{3}$ each).
\label{fig:coeffgam}}}
\end{center}
\end{figure}

Figure \ref{fig:coeffgam} shows the correlation coefficient, the veto coefficient, the energy rejection and the total rejection as a function of
the $\gamma$-ray energy. A good correlation is achieved for high energy $\gamma$-rays ($E_{\gamma} \geq $1 MeV), whereas low
energy $\gamma$-rays are mainly rejected by the veto. In fact, 80 keV X-rays undergo photoelectric effect in
the copper layer ($\sigma_{phot} \simeq 10^{4}$barn)
surrounding the cell; the scaterred electrons may escape the copper layer and leave a few keV in the cell. This will mainly happen
in the outermost cells.\\
Figure \ref{fig:rejall} presents the total rejection as a function of the $\gamma$-ray energy. It can be concluded that an \hetrois
matrix provides a rejection ranging between 10 and 1000, depending on the $\gamma$-ray energy. It must be pointed out that this is
the rejection power of the matrix itself. For instance, 90\% of X-rays will be rejected by the matrix, but the flux of
such particles will be reduced substantially by an inner and outer copper shielding.
\subsubsection{Rejection against neutron background.}
\begin{figure}
\begin{center}
\includegraphics[scale=0.55]{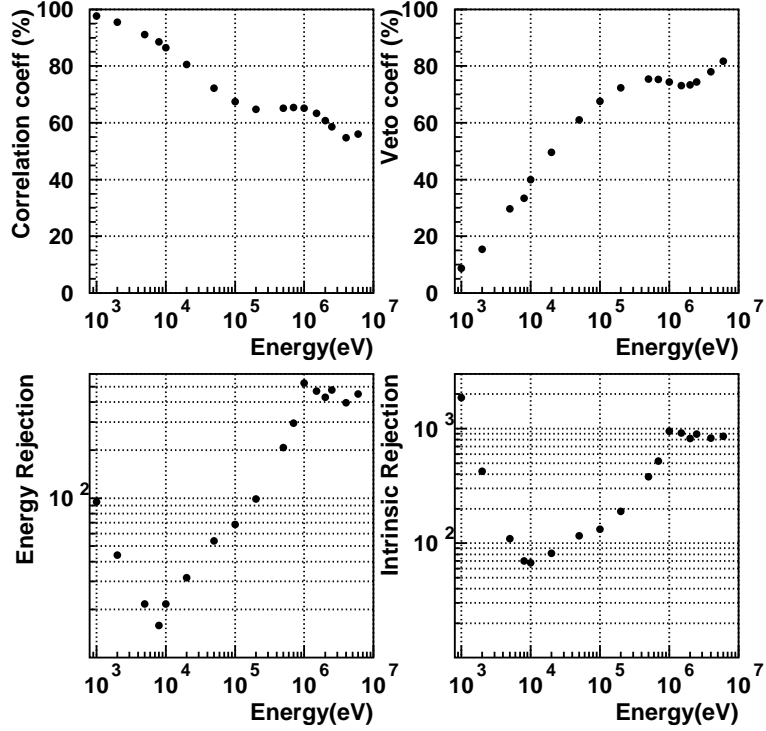}
{\noindent
\caption{Neutrons interacting in MACHe3 : The four matrix parameters, defined in sec. \ref{simumatrix}, as a function of the neutron energy, for a matrix of 1000 cells (125 $cm^{3}$ each).
\label{fig:coeffneut}}}
\end{center}
\end{figure}

Figure \ref{fig:coeffneut} presents the four matrix parameters as a function of the neutron energy. A correlation better than 70 \% is 
achieved for neutrons
of energy greater than 100 keV (fig. \ref{fig:coeffneut}, upper left), while low energy neutrons are mainly rejected by the veto. Indeed, 60 \% of 10 keV neutrons are captured in
the first layer (fig. \ref{fig:coeffneut}, upper right). The energy measurement constitute an efficient selection for low energy 
neutron (R$_{ener}\simeq$ 100 for 1 keV neutrons) and for
fast neutrons (R$_{ener}\simeq$ 1000 for 1 MeV neutrons). As expected 10 keV neutrons have the worst energy rejection (R$_{ener}\simeq$ 15).\\
The total rejection\footnote{This coefficient takes into account the interaction probability and will be used to evaluate the 
false event rate.} (ratio between number of incoming particles and number of false \neut events), shown on figure \ref{fig:rejall},
indicates that only one 1 keV
neutron out of 2000 may simulate a \neut event. The rejection falls down to 75 for 10 keV neutrons (mainly rejected by the veto) and is of the
order of 1000 for 1 MeV neutrons.

\noindent 
It must be pointed out that the evaluated rejection is for a "naked matrix", i.e. without taking into account any lead or paraffin shielding
 or any separation between electron and ion recoils. It represents the capability of the \hetrois matrix to reject background events by means of energy loss
measurements and correlation considerations. As a conclusion, it can be said that the \hetrois matrix presents a rejection power ranging between 75 and 2000 for
neutrons, and between 10 and 800 for $\gamma$-rays,depending on their kinetic energies.

\subsection{An evaluation of the neutron-induced false event rate.}
\begin{figure}
\begin{center}
\includegraphics[scale=0.60]{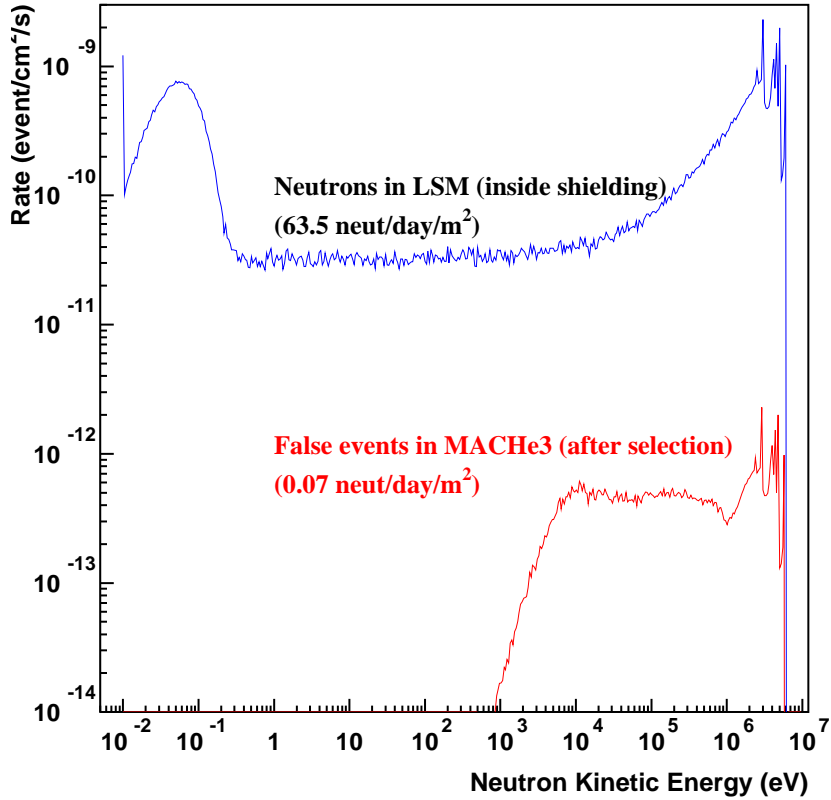}
{\noindent
\caption{The upper curve is the simulated neutron spectrum through a 30 $cm$ wide paraffin 
shielding, the measured spectrum at LSM being the input. Comparing with the measured neutron flux, this shielding allows 
an overall reduction factor of $\sim$ 50. The lowest curve represents the neutron induced false \neut rate 
in MACHe3. This spectrum has been obtained by combining the upper curve with the total rejection 
(fig. \ref{fig:rejall}), with a threshold of 500 eV. The overall counting rate, due to neutron background, is 
evaluated to be $\sim$ 0.1 day$^{-1}$ in a 1000 cells detector.
\label{fig:neutfalse}}}
\end{center}
\end{figure}

As neutrons recoiling off nuclei may easily simulate a \neut event, it is crucial to evaluate the neutron-induced false event rate.\\
In contrast to most DM detectors, MACHe3 may be sensitive to rather low energy neutrons, and its response depends strongly on their kinetic
energies. For this purpose, a simulation 
of a paraffin neutron shielding has been done, in order to evaluate the expected neutron spectrum trough this shielding.\\
The simulated device is a large ($1m\times 1m$) paraffin block ($\rho$=0.95 g.cm$^{-3}$) with a width of 30 $cm$. 
In order to be conservative, as well as keeping the calculation times short, we choose to generate the events in a direction perpendicular 
to the face of the paraffin block and considering that all neutrons crossing the block are supposed to enter the matrix volume.\\
A benchmark study has been done, to compare MCNP calculation code\cite{mcnp} and GEANT3.21. We found that these two codes give similar results, 
except for thermal neutrons (below 1 eV) for which GEANT underevaluate the flux. Again, to be conservative, we choose the one giving 
the highest flux (MCNP)\footnote{MCNP is much faster than GEANT, in this case, allowing shorter calculation times.}.\\
We have used the measured neutron spectrum \cite{Chazal:1998qn} in Laboratoire Souterrain de Modane (LSM), between 2 and 6 MeV \footnote{The thermal neutron flux, evaluated 
in \cite{Chazal:1998qn} to be (1.6$\pm0.1)\times 10^{-6} cm^{-2}s^{-1}$, will be highly suppressed by the 30 $cm$ paraffin shielding}, with an
integrated flux of $\Phi_{n}\simeq 4\times 10^{-6} cm^{-2}s^{-1}$. We found an overall neutron flux through the shielding of 5.1$\times 10^{-8} cm^{-2}s^{-1}$, with the 
neutron kinetic energy ranging between
$10^{-2}$ eV and 6 MeV (see the upper curve on fig. \ref{fig:neutfalse}).\\
Using this flux and the expected rejection factor (fig. \ref{fig:rejall}), we evaluated the false
\neut rate induced by neutron background, see fig. \ref{fig:neutfalse}. We 
found a rate of $\sim$0.1 false event per day through the
1.5$m^{2}$ surface detector (1000 cells of 125 $cm^{3}$). Even with such a conservative approach, this contamination is much lower than the expected \neut rate (of the order of 
$\sim$1 day$^{-1}$
in a detector of this size \cite{next}). 
\subsection{An evaluation of the muon-induced false event rate.}
The muon background flux in an underground laboratory (Gran Sasso) has been measured by \cite{muons}. They found a mean flux of 
$\Phi_{\mu}=2.3\times 10^{-4} m^{-2}s^{-1}$
for an average kinetic energy $<\!$$E$$>$=200 GeV.\\
An evaluation of the $\mu$-induced event rate has been done. The same procedure as above (see section \ref{simumatrix}) has been used,   
without paraffin shielding; i.e. the events are
generated in a direction perpendicular to one of the matrix faces. This is a conservative approach because the worst case is in which muons are  
passing in between 2 cell layers. As expected, most of the $\mu$-events interact in all the crossed cells (75\%
interact in 10 cells, with an average energy left of $\sim$ 1 MeV). The correlation coefficient is C$_{geo}
\simeq 2.1\%$ (meaning 97.9$\%$ of $\mu$-events are rejected), with an energy rejection R$_{ener} \simeq 40$, leading to an overall rejection 
of R$\simeq 2100$. This lead to a
$\mu$-induced false \neut rate of the order of 0.0095 day$^{-1}$m$^{-2}$, which is more than two orders of magnitude below the
expected \neut rate. The layers may be shifted, thus allowing a much higher rejection against muon background
events.
\subsection{\gam background.}
As shown in section \ref{gamrejsec}, a high granularity superfluid \hetrois detector provides an intrinsic rejection ranging 
between 10 and 800 for $\gamma$-rays, depending on their kinetic energies. This selection, based on the correlation among the cells 
and energy loss measurement, may be improved by adding a discrimination between
recoils and electrons. Different experimental approaches should be tested. A complete study of an inner and outer cryostat shielding is also
needed, as well as an evaluation of natural radioactivity of materials. Nevertheless, this simulation indicates that an important intrinsic rejection can be achieved.

\section{Conclusion}
In this prospective paper, we have demonstrate that a large matrix ($\sim$ 1000 cells) of large cells (125 $cm^{3}$) is the
preferred design for a superfluid \hetrois detector searching for DM, as far as background rejection is concerned. 
An experimental work needs to be done to demonstrate the possibility to use such a large volume of superfluid $ ^{3}{\mathrm{He}}$-B at ultra-low temperatures. This work has evaluated the background
rejection of a high granularity superfluid \hetrois detector for a large range of kinetic energies,
both for neutrons and $\gamma$-rays. It has been shown that, by means of
correlation among the cells and energy loss measurement, a high rejection may be obtained for $\gamma$-ray, neutron and muon background. Using the
measured muon and neutron flux in an underground laboratory, we have evaluated the contamination to be one order of 
magnitude (two orders
for muons) less than the expected \neut rate. For background rejection purpose, the main advantage of a superfluid \hetrois detector is to present a high 
rejection against neutron background, mainly because of the
high capture cross-section at low energy. As neutrons interact {\it a priori} like \neutt, they are the
ultimate background noise for DM detectors.\\

\textbf{Acknowledgements}\\
The authors are grateful to D. Kerdraon and L. Perrot for the help concerning the use of MCNP calculation code, and also
F. Ohlsson-Malek for the fruitful discussions on the GEANT code.

\newpage
\listoffigures

%
\newpage
\newpage
\newpage
\newpage

\newpage

\newpage
\newpage
\newpage

\end{document}